\documentclass[twocolumn,showpacs,preprintnumbers,amsmath,amssymb,APSl,prd,nofootinbib,superscriptaddress]{revtex4-1}

\usepackage{bm}
\usepackage{mathrsfs}
\usepackage{xcolor,color,graphicx,graphics}
\usepackage[all]{xy}
\usepackage{epsfig,subfigure}
\usepackage{latexsym,amssymb,amsmath,amsfonts} 
\usepackage[english]{babel} 
\usepackage[OT1]{fontenc}
\usepackage[latin1]{inputenc}
\usepackage{makeidx}
\usepackage{hyperref}
\usepackage{color,graphicx,graphics,wrapfig,epsf}

\definecolor{red}{rgb}{1,0,0}

\def\+{^\dagger}

\def\<{\leftarrow}
\def\>{\rightarrow}

\def\({\left(}
\def\){\right)}

\newcommand{\bi}{\begin{itemize}} 				\newcommand{\ei}{\end{itemize}}
\newcommand{\benu}{\begin{enumerate}} 		\newcommand{\enu}{\end{enumerate}}
\newcommand{\bd}{\begin{dinglist}{0}}     \newcommand{\ed}{\end{dinglist}}
\newcommand{\bfig}{\begin{figure}[htbp]}  \newcommand{\efig}{\end{figure}}
        			
\newcommand{\bc}{\begin{center}} 				  \newcommand{\ec}{\end{center}}
\newcommand{\be}{\begin{equation}} 				\newcommand{\ee}{\end{equation}}
\newcommand{\bsub}{\begin{subequations}}  \newcommand{\esub}{\end{subequations}}
\newcommand{\ben}{\begin{eqnarray}} 			\newcommand{\een}{\end{eqnarray}}
\newcommand{\ba}[1]{\begin{array}{#1}} 		\newcommand{\ea}{\end{array}}
\newcommand{\bea}{\begin{equation}\begin{array}{rcl}}
\newcommand{\eea}{\end{array}\end{equation}}

\begin{document}
\title{Minimum main sequence mass in quadratic Palatini $f(\mathcal{R})$ gravity}

\author{Gonzalo J. Olmo} \email{gonzalo.olmo@uv.es}
\affiliation{Departamento de F\'{i}sica Te\'{o}rica and IFIC, Centro Mixto Universidad de Valencia - CSIC.
Universidad de Valencia, Burjassot-46100, Valencia, Spain}
\affiliation{Departamento de F\'isica, Universidade Federal da
Para\'\i ba, 58051-900 Jo\~ao Pessoa, Para\'\i ba, Brazil}
\author{Diego Rubiera-Garcia} \email{drubiera@ucm.es}
\affiliation{Departamento de F\'isica Te\'orica and IPARCOS, Universidad Complutense de Madrid, E-28040 Madrid, Spain}
\author{Aneta  Wojnar}
\email{aneta.wojnar@cosmo-ufes.org}
\affiliation{N{\'u}cleo Cosmo-ufes \& PPGCosmo, Universidade Federal do Esp{\'i}rito Santo,
29075-910, Vit{\'o}ria, ES, Brazil}

\date{\today}

\begin{abstract}
General Relativity yields an analytical prediction of a minimum required mass of roughly $\sim 0.08-0.09 M_{\odot}$ for a star to stably burn sufficient hydrogen to fully compensate photospheric losses and, therefore, to belong to the main sequence. Those objects below this threshold (brown dwarfs)  eventually cool down without any chance to stabilize their internal temperature. In this work we consider quadratic Palatini $f(\mathcal{R})$ gravity and show that the corresponding newtonian hydrostatic equilibrium equation contains a new term whose effect is to introduce a weakening/strenghtening of the gravitational interaction inside astrophysical bodies. This fact modifies the General Relativity prediction for this minimum main sequence mass. Through a crude analytical modelling we use this result in order to constraint a combination of the quadratic $f(\mathcal{R})$ gravity parameter and the central density according to astrophysical observations.
\end{abstract}

\maketitle

\section{Introduction}
General Relativity (GR) is undoubtedly a successful theory of the gravitational interaction. It has been confirmed by a large array of observations/experiments \cite{Will:2014kxa}, and recently further supported by the remarkable finding of gravitational waves out of binary mergers \cite{TheLIGOScientific:2017qsa,TheLIGOScientific:2017first} (see \cite{Barack:2018yly} for a review), and by the imaging of the shadow of the supermassive black hole of M87 \cite{Akiyama:2019cqa}. However, GR also faces a number of shortcomings, including the non-detection of dark matter/energy sources needed for the consistence of the cosmological concordance model \cite{Copeland:2006wr,Nojiri:2006ri,Capozziello:2007ec,Carroll:2004de}, or the long-lasting issue of its ultraviolet completion \cite{ParTom,BirDav} and the troubles with space-time singularities \cite{Senovilla:2014gza}. Those shortcomings can be addressed via extensions of GR under the paradigm of the so-called modified theories of gravity. The later can be realized, for instance, by extending the Einstein-Hilbert action to be some more general function of curvature scalars \cite{DeFelice:2010aj}, by adding minimally or non-minimally coupled scalar fields \cite{brans, Bergmann}, by including additional geometric ingredients \cite{BeltranJimenez:2019tjy},  or by treating the physical constants as dynamical quantities  \cite{Dabrowski:2012eb, Leszczynska:2014xba}. Many such theories are now heavily constrained by gravitational wave observations \cite{Ezquiaga:2017ekz,Baker:2017hug,Creminelli:2017sry,Langlois:2017dyl,Sakstein:2017xjx,Lombriser:2015sxa}.

Astrophysical sources provide also valuable information and constraints on GR  and its extensions \cite{Berti:2015itd}. For instance, the observations of several neutron stars of two solar masses \cite{lina,as,craw} pose a challenge to our theories of the nuclear and gravitational interactions under the most extreme accessible conditions due to the unavoidable extrapolation of the form of the equation of state above the nuclear saturation density at the center of such stars.

There exists another family of objects known as brown dwarfs, which correspond to sub-stellar objects found in the lower edge of the main sequence of the Hertzsprung-Russell diagram. Brown dwarfs have central densities $\rho_c \sim 10-10^{3}$gr/cm$^3$ and radius $\sim 0.1R_{\odot}$, are composed predominantly of metallic hydrogen and helium but, not being massive enough, their contraction is halted at the onset of electronic degeneracy pressure before being able to ignite sufficient nuclear fuel to fully compensate their surface energy losses, thus eventually cooling down into oblivion. However, unlike neutron stars, whose actual composition and dynamics at different depths is a matter of active debate, brown dwarfs turn out to have rather generic and robust properties. They exhibit weak variations in their metallicity and opacity, are chemically homogeneous almost everywhere (apart from their photospheres), and can be treated as static objects because of their negligible chemical evolution (for a broad description of brown dwarfs, see \cite{Burrows:1992fg}). Thus, the weak dependence of brown dwarfs on non-gravitational physics turns them into excellent laboratories to test the predictions of modified theories of gravity.

Though this brown dwarf family encompasses a large variety of objects, here our focus will be their high-mass branch ($ \gtrsim 0.07M_{\odot}$), for which an analytic (though crude) modelling of their structural properties can be implemented. Via such a modelling one can analyze the conditions allowing thermonuclear ignition before the fluid becomes degenerate in a self-limiting process known as the minimum mass threshold required for stable hydrogen burning or, in other words, the required minimum main sequence mass (MMSM). For this process GR provides a bound of roughly $\sim 0.08-0.09 M_{\odot}$ (depending on different elements of the modelling), therefore, ``stars" below this threshold cannot hope to join the main sequence. Since dwarf stars are well described by simple polytropic equations of state, they turn out to be particularly suitable to constrain any theory of gravity predicting a modification of the hydrostatic equilibrium equation of GR inside astrophysical bodies, in particular, via its corresponding prediction for the MMSM and its compatibility with observations of the lowest-mass main sequence stars ever observed. The viability of this procedure was illustrated by Sakstein \cite{Sakstein:2015zoa,sak} for certain classes of scalar-tensor theories  where the hydrostatic equilibrium equation picks up a new term, allowing to put stringent constraints upon the underlying theory\footnote{A more recently work on this issue is the one of \cite{Crisostomi:2019yfo}, where constraints from MMSM for Degenerate Higher-Order Scalar Tensor (DHOST) theories are also found.}.

In this work we focus on  $f(\mathcal{R})$ theories of gravity formulated in metric-affine (or Palatini) spaces, where the metric and the affine connection are regarded as independent entities \cite{Olmo:2011uz}. There are several advantages of this approach. Indeed, Palatini theories of gravity modify the GR gravitational dynamics via non-linearities induced by the affine connection, which appear on the right-hand-side of the field equations as extra matter contributions. This clearly distinguishes these theories from other approaches where new propagating degrees of freedom arise. As the vacuum equations of these theories (as well as their solutions) reduce to those of GR with a cosmological constant term, they are consistent with the fact that orbital motions of binary systems should be in good agreement with vacuum GR. However, inside astrophysical bodies these theories modify the Tolman-Oppenheimer-Volkoff (TOV) equations of hydrostatic equilibrium by means of new (energy) density-dependent contributions. All these features make these theories suitable to test deviations on the MMSM with respect to the GR result. Here we consider the (perhaps) simplest member of the Palatini $f(\mathcal{R})$ family, namely, the quadratic one, for which physically compelling results on black holes and the avoidance of space-time singularities have been obtained by some of us \cite{Bambi:2015zch,Bejarano:2017fgz}. For this particular theory, the corresponding TOV equations can be solved by going to the Einstein frame. Since observations have narrowed down the minimum threshold of very-low mass stars to $M \gtrsim 0.0930 \pm 0.0008 M_{\odot}$ corresponding to the M-dwarf star G1 866C \cite{Segransan:2000jq}, comparison of the predictions of our model with this observation allows us to constrain a combination of the parameter of quadratic $f(\mathcal{R})$ gravity and the star's central density.

This work is organized as follows: in Sec.\ref{secP} we introduce quadratic Palatini $f(\mathcal{R})$ gravity, cast its field equations for perfect fluids into the modified stellar hydrostatic equilibrium (TOV) equations, and obtain the non-relativistic limit of such equations (generalized Lane-Emden equation). From this equation, the relevant physics of brown dwarfs for the MMSM is analyzed in Sec. \ref{sec:MMHB}, whose main finding is an expression for it involving the quadratic gravity parameter and the star's central density. Sec. \ref{sec:con} contains a discussion of the results obtained and the limitations of our approach as well as some future perspectives.

\section{Stellar equilibrium equations in Palatini $f(\mathcal{R})$ gravity}\label{secP}

In the Palatini formulation of gravitational theories the geometry and its dynamics is encapsulated into two independent structures, a class of Lorentzian metrics $g_{\mu\nu}$ and an affine connection $\Gamma \equiv \Gamma_{\mu\nu}^{\lambda}$. Here we are considering the simplest extension of GR within this formulation, namely, $f(\mathcal{R})$ theories, whose action is given by
\begin{equation} \label{eq:action}
\mathcal{S}=\mathcal{S}_{G}+\mathcal{S}_{m}=\frac{1}{2\kappa}\int d^4x \sqrt{-g}f(\mathcal{R})+\mathcal{S}_{m}(g_{\mu\nu},\psi_m) \ ,
\end{equation}
where $\kappa \equiv -8\pi G/c^4$ is Newton's constant (from now on $G=c=1$), $g$ is the determinant of the space-time metric, the affine connection is built in the Ricci tensor $\mathcal{R}_{\mu\nu}(\Gamma)\equiv {\mathcal{R}^\alpha}_{\mu\alpha\nu}$, and $f(\mathcal{R})$ is some function of the Ricci scalar $\mathcal{R}=g^{\mu\nu} \mathcal{R}_{\mu\nu}(\Gamma)$. As for the matter action, $\mathcal{S}_{m}$, it is assumed to be minimally coupled, with $\psi_m$ representing collectively the matter fields. Variation of the action (\ref{eq:action}) with respect to $g_{\mu\nu}$ and $\Gamma$ yields two systems of equations:
\begin{align}
   f_\mathcal{R} \mathcal{R}_{\mu\nu}-\frac{1}{2}f(\mathcal{R})g_{\mu\nu}=&\kappa T_{\mu\nu},\label{structural}\\
   \nabla^{\Gamma}_\beta(\sqrt{-g}f_\mathcal{R}g^{\mu\nu})=&0 \ ,\label{con}
\end{align}
where $f_\mathcal{R} \equiv df/d\mathcal{R}$ and $T_{\mu\nu}=\frac{2}{\sqrt{-g}} \frac{\delta \mathcal{S}_m}{\delta g^{\mu\nu}}$ is the energy momentum tensor
of the matter fields. Let us first note that tracing over the system of equations (\ref{structural}) one finds that $\mathcal{R}f_\mathcal{R}-2f=\kappa T$, which is an algebraic equation telling us that the curvature scalar can be removed in favour of the matter sources (via the trace $T$ of the energy-momentum tensor). In turn, this allows to interpret Eq.(\ref{con}) as the standard compatibility condition of the independent connection $\Gamma$ with another rank-two tensor $h_{\mu\nu}$, conformally related to the space-time metric $g_{\mu\nu}$ as
\begin{equation} \label{eq:conft}
    h_{\mu\nu}=f_\mathcal{R} g_{\mu\nu} \ .
\end{equation}
In other words, $\Gamma$ is Levi-Civita of $h_{\mu\nu}$ while $g_{\mu\nu}$ is obtained via the conformal transformation (\ref{eq:conft})
with $f_\mathcal{R}\equiv f_\mathcal{R}(T)$ fully determined by the matter sources once some functional form $f(\mathcal{R})$ is given. In particular,
when $f(\mathcal{R})=\mathcal{R}-2\Lambda$, then $f_R$ trivializes, the independent connection becomes Levi-Civita of $g_{\mu\nu}$, and the field
equations (\ref{structural}) reduce to the standard Einstein equations (with cosmological constant $\Lambda$) of GR. Similarly, in vacuum, ${T_\mu}^{\nu}=0$, or
for traceless matter, $T=0$, one finds that $f_R$ also trivializes independently of its form and one recovers the GR dynamics as well. This implies that Palatini $f(\mathcal{R})$ models
are ghost-free, propagating only the two polarizations\footnote{It is well known that the metric formulation of $f(R)$ theories is equivalent to Brans-Dicke theories with  $\omega=0$, while the Palatini version leads instead to $\omega=-3/2$. Given that the scalar field equation in Brans-Dicke theory is of the form $(3+2\omega)\Box \phi +\phi V_\phi - 2V=\kappa T$, it follows that in the $\omega=-3/2$ case there is no dynamics for the scalar and, therefore, the resulting theory has exactly the same number of dynamical degrees of freedom as GR. A first discussion of this point can be found in \cite{Olmo:2005zr,Olmo:2005hc}  while a Hamiltonian analysis of the degrees of freedom was provided in \cite{Olmo:2011fh}. } of the gravitational field (gravitational waves) travelling at the speed of light.

To handle the field equations (\ref{structural}) in a more convenient way for the sake of the problem considered here, we use the fact that they can be rewritten  in the terms of the
conformal metric $h_{\mu\nu}$ \cite{BSS,SSB} and the scalar field $\Phi \equiv f_\mathcal{R}$ as
 \begin{eqnarray}
	 \bar{\mathcal{R}}_{\mu\nu} - \frac{1}{2} h_{\mu\nu} \bar{\mathcal{R} } &=&\kappa \bar T_{\mu\nu}-\frac12 h_{\mu\nu} \frac{U(\Phi)}{\Phi^2} \ , \label{EOM_P1} \\
	  2 U(\Phi) - \Phi U_\Phi &=& \kappa  T
	\label{EOM_scalar_field_P1} \ ,
	\end{eqnarray}
where $\Phi\equiv f_\mathcal{R}$, $U(\Phi)\equiv(\mathcal{R}\Phi-f(\mathcal{R}))$ and $\bar T_{\mu\nu}\equiv\Phi^{-1}T_{\mu\nu}$. One must also bear in mind that $\mathcal{R}_{\mu\nu}=\bar{\mathcal{R}}_{\mu\nu}, \bar{\mathcal{R}}= h^{\mu\nu}\bar{\mathcal{R}}_{\mu\nu}=\Phi^{-1} \mathcal{R}$ and $h_{\mu\nu}\bar{\mathcal{R}}=\ g_{\mu\nu} \mathcal{R}$. The system (\ref{EOM_P1})- (\ref{EOM_scalar_field_P1}) corresponds to the field equations of an Einstein-like theory in which the metric  $h_{\mu\nu }$ is sourced by non-linear terms associated to the matter fields. The non-linearities enter through the scalar quantity $\Phi$, which is algebraically determined by the matter sources via Eq.(\ref{EOM_scalar_field_P1}), which implies $\Phi=\Phi(T)$.

Using the above definitions and properties let us now assume as our matter source a perfect fluid with energy-momentum tensor
\begin{equation}
{T_\mu}^{\nu}=(\rho+p)u_{\mu}u^{\nu} +p\delta_{\mu}^{\nu} \ ,
\end{equation}
where the unit vector $u_{\mu}u^{\mu}=-1$, while $\rho$ and $p$ are the energy density and pressure of the fluid, respectively. It has been shown in \cite{aneta, anet} that, for a static, spherically symmetric line element,
the generalized TOV equations of stellar hydrostatic equilibrium for this theory read
\begin{eqnarray}
  \left(\frac{\Pi}{\Phi({r})^2}\right)'&=&-\frac{GA\mathcal{M}}{r^2}\left(\frac{Q+\Pi}{\Phi({r})^2}\right)
  \left(1+\frac{4\pi r^3\frac{\Pi}{\Phi({r})^2}}{\mathcal{M}}\right)\label{tov_kon}\\
A&=&1-\frac{2G \mathcal{M}(r)}{r} \\
\mathcal{M}(r)&=& \int^r_0 4\pi \tilde{r}^2\frac{Q(\tilde{r})}{\Phi(\tilde{r})^2} d\tilde{r} \ ,\label{mass}
\end{eqnarray}
where primes stand for radial derivatives, and the generalized energy density $Q$ and pressure $\Pi$ are defined as
\begin{subequations}\label{defq}
 \begin{equation}
   \bar{Q}=\bar{\rho}+\frac{\bar{U}}{2\kappa c^2}=\frac{\rho}{\Phi^2}+\frac{U}{2\kappa c^2\Phi^2}=\frac{Q}{\Phi^2} \ ,
 \end{equation}
\begin{equation}
  \bar{\Pi}=\bar{p}-\frac{\bar{U}}{2\kappa}=\frac{p}{\Phi^2}-\frac{U}{2\kappa\Phi^2}=\frac{\Pi}{\Phi^2} \ .
\end{equation}
\end{subequations}
Recall that $\bar{U}$ and $\Phi$ depend on the choice of the gravitational model one is interested in. It is readily seen that, in the GR limit, $\Phi=1,\bar{U}=0$, one recovers the standard TOV equations.

\subsection{Generalized Lane-Emden equation}

In what follows we shall focus on the quadratic model
\begin{equation} \label{eq:staro}
f(\mathcal{R})= \mathcal{R}+\beta \mathcal{R}^2  \ ,
\end{equation}
first introduced by Starobinsky \cite{Starobinsky:1980te} in the context of inflation. This model should capture in an effective way relevant contributions from higher-order modifications of GR. Though other kind of quadratic terms are certainly possible, such as $\mathcal{R}_{\mu\nu}\mathcal{R}^{\mu\nu}$ corrections\footnote{It should be noted that Ricci-squared corrections in the metric formulation generically lead to ghost-like instabilities \cite{Stelle:1977ry,Deruelle:2009zk}. In the Palatini version that we are considering here, however, the situation is completely different. In fact, the Palatini version of the so-called Ricci-Based Gravity theories, in which the Lagrangian is an arbitrary function of the metric and the symmetric part of the Ricci tensor, always leads to second-order field equations which recover GR in vacuum. For details see  \cite{Afonso:2017bxr}.} , the technical simplicity of the $f(\mathcal{R})$ model over any other extension justifies our consideration of this model.

To carry out the analysis of the MMSM one takes advantage of the fact that brown stars are non-relativistic objects \cite{Burrows:1992fg}, and in such a case the generalized TOV equations  (\ref{tov_kon}) can be greatly simplified. In particular, the relation between the energy density and the pressure needed to close the TOV equations is typically assumed to be of polytropic type.  Though in our analysis we will rely on this standard description, one should bear in mind that the polytropic approximation is useful as far as it provides a rough idea of the structural properties of the object, such as the scale of its total mass and size. For realistic discussions of the observational features of stellar objects, however, more realistic and accurate descriptions beyond the polytropic simplification are needed. In fact, the physics of stellar atmospheres is complex and depends intimately on their temperature and composition, which are not accounted for by polytropes in any way. For this reason, though it has been claimed that polytropes pose severe constraints on Palatini $f(\mathcal{R})$ and other metric-affine theories, see e.g. \cite{Pani:2012qd,Barausse:2007pn}, our view is that such conclusions are just the result of an extreme mathematical simplification and idealization of the astrophysical problem  \cite{Olmo:2008pv}, having no actual impact on realistic physical scenarios, where metallicity, electrostatic effects, radiation fluxes, and other fine details  demand non-polytropic descriptions and substantially affect how the geometry transits towards the (non-empty) external environment (see \cite{BeltranJimenez:2017doy,Olmo:2011uz} for further discussions on these topics).

Thus, considering a polytropic equation of state (EoS)
\begin{equation} \label{eq:EoS}
p=K \rho^{\frac{n+1}{n}} \ ,
\end{equation}
where $K$ is the polytropic constant, and $n$ the polytropic index, and introducing the following redefinitions
\begin{equation}
r=r_c \bar{\xi}, \rho=\rho_c \theta^n, p=p_c \theta^{n+1}, r_c^2=\frac{(n+1)}{4\pi G \rho_c^2} p_c
\end{equation}
it was shown in \cite{aneta2} that, in the Einstein frame, the generalized Lane-Emden equation for the quadratic model (\ref{eq:staro}) is given by
\begin{equation}\label{LEo}
\theta^n + \frac{1}{\bar{\xi}}\frac{d^2}{d\bar{\xi}^2}\left(\left[1+\frac{2\alpha}{n+1}\theta^n\right]\bar{\xi}\theta\right)=0,
\end{equation}
where $\alpha=\kappa c^2\beta\rho_c$, with $\rho_c$ being the star's central density. Other generalizations of the Lane-Emden equation for scalar-tensor theories can be found in \cite{Koyama:2015oma,Saito:2015fza}. The above equation picks up a new term as compared to the Lane-Emden one of GR \cite{GlenBook}, which is recovered in the limit $\alpha \to 0$. As high-mass brown dwarfs contract along a Hayashi track they are nearly fully convective and can be well approximated by a polytropic EoS  (\ref{eq:EoS}) with $n=3/2$ \cite{KippBook} (and with $K$ a function of the specific entropy), and this is the value we shall take hereafter. Let us note that, in general, the generalized Lane-Emden equation in Palatini gravity, as well as in some other modified theories of gravity \cite{Capozziello:2011nr} is not scale-invariant because the parameter $\alpha$ depends on the central energy density $\rho_c$.

For our analysis it is more useful to cast the generalized Lane-Emden equation (\ref{LEo}) in the Jordan frame. Performing the conformal
transformation $\bar{\xi}^2=\Phi\xi^2$, where for our model (\ref{eq:staro}) we have $\Phi=1+2\alpha\theta^n$, the above equation is written as
\begin{equation}\label{LE}
\xi^2\theta^n\Phi^{3/2}+\displaystyle\frac{1}{1+\frac{\xi\Phi_{\xi}}{2\Phi}}\displaystyle\frac{d}{d\xi}\left(\displaystyle\frac{\xi^2\Phi^{3/2}}{1+ \frac{\xi\Phi_{\xi}}{2\Phi}}\displaystyle\frac{d\theta}{d\xi} \right)=0 \ .
\end{equation}
Though this equation is seemingly more involved than the one in the Einstein frame, it has some advantages, as shall be seen at once. As usual, imposing boundary conditions $\theta(0)=1$ and $\theta'(0)=0$, once a solution $\theta(\xi)$ to (\ref{LE}) is found, the first zero $\xi_R$ of $\theta(\xi_R)=0$ allows one to compute the star's mass, radius, central density, and temperature via the following expressions
\begin{align}
 M&=4\pi r_c^3\rho_c\omega_n,\label{masss}\\
 R&=\gamma_n\left(\frac{K}{G}\right)^\frac{n}{3-n}M^\frac{n-1}{n-3}\xi_R \label{radiuss},\\
 \rho_c&=\delta_n\left(\frac{3M}{4\pi R^3}\right) \label{rho0s} ,\\
 T&=\frac{K\mu}{k_B}\rho_c^\frac{1}{n}\theta_n \label{temps},
\end{align}
where $k_B$ is Boltzmann's constant and $\mu$ the mean molecular weight, and we have introduced the set of constants \cite{artur}
\begin{align}
 \omega_n&=-\frac{\xi^2\Phi^\frac{3}{2}}{1+\frac{1}{2}\xi\frac{\Phi_\xi}{\Phi}}\frac{d\theta}{d\xi}\mid_{\xi=\xi_R},\label{omega}\\
  \gamma_n&=(4\pi)^\frac{1}{n-3}(n+1)^\frac{n}{3-n}\omega_n^\frac{n-1}{3-n}\xi_R,\label{gamma}\\
 \delta_n&=-\frac{\xi_R}{3\frac{\Phi^{-\frac{1}{2}}}{1+\frac{1}{2}\xi\frac{\Phi_\xi}{\Phi}}\frac{d\theta}{d\xi}\mid_{\xi=\xi_R}} \ . \label{delta}
\end{align}
Let us note that the constants (\ref{omega}) and (\ref{delta}) differ from their GR forms in that new $\Phi$-dependent terms have been picked up. The reason for this is that $r$-coordinate in the mass function (\ref{mass}) is the one coming from the conformal metric $h_{\mu\nu}$ due to the use of Jordan frame. In GR, $\Phi_{\xi}=0$ and these constants recover their standard meanings.

From the generalized Lane-Emden equation (\ref{LE}), one finds that its solution near the center behaves as
\begin{equation}\label{origin}
 \theta(\xi \approx 0) =1-\frac{\xi^2}{6} \ ,
\end{equation}
which is the same result as in GR, and can be well approximated for the purpose of computing the MMSM \cite{FowHoy} as
\begin{equation}\label{sol1}
 \theta(\xi \approx 0)=\text{exp}\left[-\frac{\xi^2}{6}\right].
\end{equation}
These are the main elements we need from the generalized Lane-Emden equation (\ref{LE}) in order to carry out the determination of the MMSM in quadratic Palatini $f(\mathcal{R})$ gravity in next section.

\section{Minimum main sequence mass} \label{sec:MMHB}

Our analysis of the MMSM for quadratic Palatini $f(\mathcal{R})$ gravity will now parallel the one carried out by Burrows and Liebert \cite{Burrows:1992fg} for GR and, more recently, the one by Sakstein \cite{Sakstein:2015zoa,sak} for certain scalar-tensor theories. The starting point of this analysis is to note that the thermonuclear ignition is powered by three main chain reactions: $p + p \to d + e^{+}+\nu_e, p +e^{-}+p \to d +\nu_e, p+d \to {^{3}}H_e + \gamma$, where the first one is a slow process acting as the bottle-neck behind the MMSM bound. The energy generation rate per unit mass of this process can be well approximated by the power law form \cite{Burrows:1992fg}
\begin{equation} \label{eq:pp}
\dot{\epsilon}_{pp}= \dot{\epsilon}_c \left(\frac{T}{T_c}\right)^s \left(\frac{\rho}{\rho_c} \right)^{u-1} \ ,
\end{equation}
where $T_c$ and $\rho_c$ are the central temperature and density, respectively, and the two exponents can be approximated as  $s \approx 6.31$ and $u \approx 2.28$. For an assumed hydrogen fraction of $75\%$ in a high-mass brown dwarf the number of baryons per electron can be fixed to $\mu_e \approx 1.143$, which yields the value of the constant
\begin{equation} \label{eq:e0}
 \dot{\epsilon}_c=\epsilon_0T_c^s\rho_c^{u-1} \ ,
\end{equation}
where   $\dot{\epsilon}_0\approx 3.4\times10^{-9}$ ergs g$^{-1}$s$^{-1}$. Next, the polytropic constant $K$ appearing in the polytropic EoS can be approximated by an expression valid at both the low-temperature and high-temperature regimes (but not in between) of the brown dwarf  as  \cite{Burrows:1992fg}
\begin{equation}
 K=\frac{(3\pi^2)^\frac{2}{3}\hbar^2}{5m_em_H^\frac{5}{3}\mu_e^\frac{5}{3}}\left(1+\frac{\alpha_d}{\eta}\right).
\end{equation}
where $m_e$ is the electron mass, $m_H$ the mass of atomic Hydrogen, the constant  $\alpha_d\equiv\frac{5\mu_e}{2\mu}\approx4.82$,  while the quantity
\begin{equation}
\eta \equiv \frac{\mu_F}{\kappa_B T} \ ,
\end{equation}
where $\mu_F$ is Fermi energy, measures the degree of the degeneracy electron pressure of the star ($\eta \gg 1$ for fully degenerate gas and $\eta \ll 1$ for ideal gas law).
Thus,  for $n=3/2$, from (\ref{radiuss}), we get the stellar radius as
\begin{equation}\label{rad1}
 R=\frac{(3\pi^2)^\frac{2}{3}\hbar^2}{5Gm_em_H^\frac{5}{3}\mu_e^\frac{5}{3}}\left(1+\frac{\alpha_d}{\eta}\right)\gamma_{3/2}M^{-1/3} \ ,
\end{equation}
which one uses to obtain the core density in (\ref{rho0s}) as
\begin{equation}\label{cored}
 \rho_c=\frac{125G^3m_e^3m_H^5\mu^5_e}{12\pi^5\hbar^6}\frac{\delta_{3/2}}{\gamma^3_{3/2}}M^2\left(1+\frac{\alpha_d}{\eta}\right)^{-3} \ ,
\end{equation}
while the core temperature follows from (\ref{temps})  as
\begin{equation}
 T_c= \frac{25G^2m_em_H^{8/3}\mu^{8/3}_e}{2^{7/3}\pi^2\hbar^2}\frac{\delta^\frac{2}{3}_{3/2}}{\gamma^2_{3/2}}\frac{\eta}{(\alpha_d+\eta)^2}M^\frac{4}{3} \ .
\end{equation}
From the energy generation rate formula (\ref{eq:pp}), one can integrate over the stellar volume to find the luminosity from hydrogen burning as
 \begin{equation}
  L_{HB}=4\pi r_c^3\rho_c\dot{\epsilon}_c\int^{\xi_R}_{0}\xi^2\theta^{n(u+\frac{2}{3}s)}d\xi \ .
 \end{equation}
Taking the approximate solution (\ref{sol1}) this integral can be computed as\footnote{In doing so we take into account that most of the star's mass is concentrated at the center, and that other uncertainties and approximations used in this crude modelling will be larger than those involved in the use of the approximated solution (\ref{sol1}). In particular, this implies to keep the new gravitational scale $\alpha$ small enough, which will be the case for the numerical computations employed later.}
\begin{equation} \label{eq:LHB}
 L_{HB}=\frac{3\sqrt{6\pi}}{2\omega_{3/2}(\frac{3}{2}u+s)^\frac{3}{2}}\dot{\epsilon}_c M \approx 0.079 \dot{\epsilon}_c M \ .
\end{equation}
 Let us note again that the modification from quadratic Palatini gravity appears both in the mass $M$ and in the quantity $\omega_{3/2}$, as follows from Eqs.(\ref{masss}) and (\ref{omega}), via the conformal factor $\Phi$. Therefore, inserting these formulas as well as Eq.(\ref{eq:e0}) in the luminosity formula (\ref{eq:LHB}) we find the result
\begin{equation}
 L_\text{HB}=1.53\times10^7L_\odot\frac{\delta^{5.487}_{3/2}}{\omega_{3/2}\gamma^{16.46}_{3/2}}M^{11.977}_{-1}
 \frac{\eta^{10.15}}{(\eta+\alpha_d)^{16.46}} \ ,
\end{equation}
where we have defined $M_{-1}=M/(0.1M_\odot)$ and $L_{\odot}$ is the solar luminosity.

A star burns hydrogen in a stable way when the above luminosity is equal to the luminosity at the photosphere, $L_{ph}$. The photosphere is defined at the radius for which the optical depth
\begin{equation} \label{eq:od}
 \tau(r)=\int_r^\infty \kappa_R\rho dr \ ,
\end{equation}
equals $2/3$. In this  formula, $\kappa_R$ stands for Rosseland's mean opacity. The photosphere lies indeed very close to the stellar radius and therefore we will approximate this radius as the stellar one in what follows. To keep going, let us come back to the modified hydrostatic equilibrium equation and the mass in the Newtonian limit, which read explicitly, in the Einstein frame, as
\begin{align}
 \frac{dp}{d\tilde r}&=-\frac{G\mathcal{M}\rho}{\phi\tilde r^2} \ ,\\
\mathcal{M}&=\int_0^{\tilde r}4\pi x^2\rho(x)dx \ ,
\end{align}
where $\tilde r^2=\phi r^2$ and $\phi=1+2\kappa^2 c^2\beta \rho$. Transforming back to the Jordan frame, Taylor-expanding around $\beta=0$ reduces the modified hydrostatic equilibrium equation to
\begin{equation}\label{pres}
 p'=-g\rho(1+\kappa c^2 \beta [r\rho'-3\rho]) \ ,
\end{equation}
where $g=\text{const}$ is the surface gravity, which can be approximated as
\begin{equation}\label{surf}
 g\equiv\frac{G\mathcal{M}(r)}{r^2}\sim\frac{GM}{R^2} \ .
\end{equation}
We also need to transform the mass function $\mathcal{M}(r)$. Since the transformation depends on the energy density, which on the star's surface drops to $\rho\approx 0$, we assume that also in the Jordan frame the following is true
\begin{equation}
 \mathcal{M}''=8\pi r\rho+4\pi r^2 \rho' \ ,
\end{equation}
which we use in (\ref{pres}) to find
\begin{equation}
 p'_{ph}=-g\rho\left( 1+8\beta\frac{g}{c^2 r} \right).
\end{equation}
Using the definition of the optical depth (\ref{eq:od}) we can integrate the above expression as
\begin{equation} \label{eq:pressurephoto}
 p_{ph}=\frac{2g\left( 1+8\beta\frac{g}{c^2 R} \right)}{3\kappa_R} \ ,
\end{equation}
where the radius $R$ is given by (\ref{rad1}). Let us note that one can cast this expression in terms of the GR one by defining the effective opacity
\begin{equation}
 \kappa^\text{eff}_R=\frac{\kappa_R}{\left(1+8\beta\frac{g}{c^2 R}\right)} \ ,
\end{equation}
which, for the quadratic gravity considered here, depends not only on the theory parameter, $\beta$, but also on the surface gravity and the core density via the stellar radius. This introduces a fundamental difference with other theories of gravity, such as scalar-tensor theories, where this modification only involves the theory's parameter \cite{Sakstein:2015zoa,sak}. Assuming the ideal gas law, from Eq.(\ref{eq:pressurephoto}) one can write
\begin{equation}\label{gas}
 \frac{\rho k_B T}{\mu m_H}=\frac{2g\left(1+8\beta\frac{g}{c^2 R} \right)}{3\kappa_R} \ .
\end{equation}
The surface gravity $g$, after inserting the expressions for the mass and radius (\ref{rad1}), and making explicit the numerical values of the constants, is written as
\begin{equation}
 g=\frac{3.15\times10^6}{\gamma^2_{3/2}}
M_{-1}^{5/3}\left(1+\frac{\alpha_d}{\eta}\right)^{-2} \text{cm/s}^2 \ ,
\end{equation}
while the photospheric temperature  can be obtained from matching the specific entropy of the gas and metallic phases there, which yields the result (we refer again to Barrows and Liebert \cite{Burrows:1992fg} for details)
\begin{equation}\label{temp}
 T_{ph}=1.8\times10^6\frac{\rho_{ph}^{0.42}}{\eta^{1.545}}\text{K} \ .
\end{equation}
Thus, applying those two expressions to (\ref{gas}) we find the photospheric energy density
\begin{align}
 \frac{\rho_{ph}}{\mathrm{g/cm^3}}&=
 5.28\times10^{-5}M^{1.17}_{-1}\left(\frac{1+8\beta\frac{g}{c^2 R} }{\kappa_{-2}}\right)^{0.7}\nonumber\\
 &\times\frac{\eta^{1.09}}{\gamma^{1.41}_{3/2}}\left(1+\frac{\alpha_d}{\eta}\right)^{-1.41} \ ,
\end{align}
where $\kappa_{-2}=\kappa_R/(10^{-2}\mathrm{cm^2 g^{-1}})$. Inserting $\rho_{ph}$ into the photospheric temperature (\ref{temp}) one finds
\begin{align}
 \frac{T_{ph}}{\text{K}}&=2.88\times10^4\frac{M^{0.49}_{-1}}{\gamma^{0.59}_{3/2}\eta^{1.09}}\nonumber\\
 &\times
 \left(\frac{1+8\beta\frac{g}{c^2 R} }{\kappa_{-2}}\right)^{0.296}\left(1+\frac{\alpha_d}{\eta}\right)^{-0.59} \ .
\end{align}
The stellar luminosity defined as $L_{ph}=4\pi R^2\sigma T^4_{ph}$ is found to be
\begin{align}
L_{ph}&=28.18L_\odot\frac{M^{1.305}_{-1}}{\gamma^{2.366}_{3/2}\eta^{4.351}}\nonumber\\
 &\times\left(\frac{1+8\beta\frac{g}{c^2 R} }{\kappa_{-2}}\right)^{1.183}
 \left(1+\frac{\alpha_d}{\eta}\right)^{-0.366} \ .
\end{align}
Using again the formulas (\ref{rad1}) and (\ref{surf}) to get rid of the quotient $g/R$ one writes the above expression as
\begin{align}
 L_{ph}&=28.18L_\odot\frac{M^{1.305}_{-1}}{\gamma^{2.366}_{3/2}\eta^{4.351}}\left(1+\frac{\alpha_d}{\eta}\right)^{-0.366} \nonumber\\
 &\times\left(\frac{\delta_{3/2}-1.31\alpha\left(1+\frac{\alpha_d}{\eta}\right)^4}{\delta_{3/2}\kappa_{-2}}\right)^{1.183} \ .
\end{align}
where we recall that  $\alpha \equiv  \kappa c^2  \beta \rho_c$. Finally, the MMSM is obtained from setting $L_{HB}=L_{ph}$, which yields the result
\begin{equation} \label{result}
M_{-1}^{MMSM}=0.290 \frac{\gamma_{3/2}^{1.32} \omega_{3/2}^{0.09}}{\delta_{3/2}^{0.51}} I(\eta,\alpha) \ ,
\end{equation}
where we have introduced the new function
\begin{equation} \label{eq:Ifunc}
I(\eta,\alpha)=\frac{(\alpha_d + \eta)^{1.509}}{\eta^{1.325}} \left(1-1.31\alpha\frac{\left(\frac{\alpha_d+\eta}{\eta}\right)^4}{\delta_{3/2}\kappa_{-2} }\right)^{0.111} \ .
\end{equation}
This is the master equation of this work. Since the function  $I(\eta,\alpha)$ has a minimum (in $\eta$) for every value of $\alpha$, this formula provides the MMSM for the quadratic gravity model (\ref{eq:staro}) as given by the lowest value of the mass such that Eq.(\ref{result}) is satisfied. In the next section we shall discuss the consequences for the observational viability of the parameters of this theory upon the assumptions considered so far.

\section{Discussion and conclusion} \label{sec:con}

We first point out that formula (\ref{result}) confirms the result of \cite{sak} that the
MMSM depends weakly  on the opacity $\kappa_R$, which is an element hard to model, and that we take here to be given by a reference value $\kappa_R=10^{-2}$cm$^2$g$^{-1}$ as discussed in \cite{Burrows:1992fg} (for a broader discussion of the values of the opacity in brown dwarfs depending on the density and temperature, see e.g. \cite{Freedman:2007cm}). Second, one must bear in mind that in this formula there appears the parameter $\alpha$, which contains both the quadratic gravity parameter, $\beta$, and the central density, $\rho_c$. This is a common feature of Palatini theories of gravity, in that they typically induce new energy-density dependent contributions, clearly distinguishing them from other proposals extending GR. In the present context, this element introduces novelties in discussing the constraints for the theory's parameters, as shall be seen at once. Finally we also point out that, in the problem considered here, once a zero of $\theta(\xi_R)=0$ is found by the resolution of the generalized Lane-Emden equation (\ref{LE}), this introduces modifications in the value of $\omega_n$ and $\delta_n$ but not in the conformal factor itself, $\Phi$, since for the case of $n=3/2$ (modelling high-mass brown dwarfs) this factor trivializes.  This is not so for other polytropic indices. For instance, the case $n=1$ (which can be used to model low-mass brown dwarfs) considered in \cite{artur} does introduce modifications.

\begin{figure}[t]
\centering
\includegraphics[scale=.93]{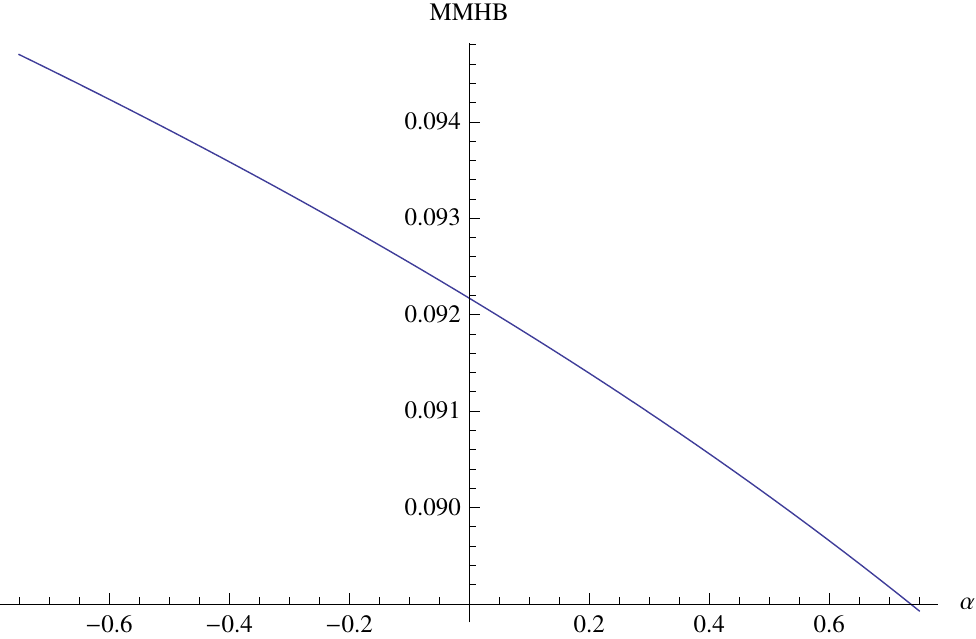}
\caption{The normalized MMSM in quadratic Palatini $f(\mathcal{R})$ gravity as a function of the parameter $\alpha= \kappa c^2\beta\rho_c$ assuming the GR values for $\omega_{3/2},\gamma_{3/2},\delta_{3/2}$, as follows from the formula (\ref{eq:MMSManalytic}).}
\label{fig.1}
\end{figure}

One might naively think that, for small values of $\alpha$, the functions (\ref{omega})-(\ref{delta}), obtained after resolution of the generalized Lane-Emden equation (\ref{LE}), should not significantly deviate from those of GR, namely, $\omega_{3/2}=2.71$, $\gamma_{3/2}=2.357$, and $\delta_{3/2}=5.991$ and that, therefore, one could safely use those values for the problem considered here. Therefore, considering that for $\alpha=0$ the function (\ref{eq:Ifunc}) peaks at $2.34$ for a degree of degeneracy  $\eta=34.7$, and using the value $\alpha_d\approx4.82$, one could write a simple analytic expression for the MMSM as a function of the parameter $\alpha$ as
\begin{equation} \label{eq:MMSManalytic}
 M_{MMSM}^{\alpha} \approx0.0922(1-0.368\alpha)^{0.111}M_\odot.
\end{equation}
For GR ($\alpha=0$) one gets $M_{MMSM}^{GR}\approx0.0922M_\odot$, which is consistent with the one recently reported in \cite{Crisostomi:2019yfo}, though slightly above the one originally obtained by Burrows and Liebert \cite{Burrows:1992fg} for the same degree of degeneracy. Plotting in Fig. \ref{fig.1} the evolution of the MMSM with (small) $\alpha$ yields a trend in which the MMSM grows (decreases) slowly with negative (positive) $\alpha$.

However, this approach turns out to be seriously flawed, as follows from a  case-by-case numerical analysis (in $\alpha$). As mentioned above, the modification to the MMSM value in quadratic Palatini $f(\mathcal{R})$ gravity does not enter only by direct contributions on the parameter $\alpha$, but also via the modifications to the value of the constants $\gamma_{3/2}$, $\omega_{3/2}$, and $\delta_{3/2}$, because they depend on the solution of the modified Lane-Emden equation. The net effect is that the MMSM in this theory is actually very sensitive to relatively mild variations in $\alpha$ via these two sources. Thus, resolving the generalized Lane-Emden equation for specific values of $\alpha$ (both positive and negative), we have presented the obtained results in Table \ref{tab}. Here we observe that, starting from the GR ($\alpha=0$) value of $M_{MMSM}^{GR}\approx 0.0922M_{\odot}$,  the actual trend of the MMSM with $\alpha$ is reversed: positive (negative) values of $\alpha$ yield larger (smaller) values of the MMSM. We find that a value as high as $\alpha=0.010$ yields $M_{MMSM} \approx 0.0933M_{\odot}$, which is already on the verge of becoming incompatible with the bound $(0.0930\pm0.0008) M_\odot$, corresponding to the mass of the M-dwarf star G1 866C \cite{Segransan:2000jq} and, therefore, values significantly above this one would be in tension with observations. The branch $\alpha <0$, on the other hand,  lowers the MMSM and, therefore, seems to be safe from any such problems.

This is how far we dare to go on constraining this parameter given the limitations and approximations involved in the crude analytical modelling employed here. Such limitations include, but are not limited to, the polytropic approximation (\ref{eq:EoS}) itself, missing information on an accurate description of the atmosphere, thermodynamical aspects, the approximations for the energy generation rates, losses on the contribution to $L_{HB}$ due to the approximation (\ref{sol1}), overestimation of a realistic value of $\eta$ \cite{Burrows:1992fg}, and so on.  As in GR, fully reliable results can only be obtained via numerical resolutions of the stellar structure equations. Indeed, since the numerical simulations tend to decrease the value of the MMSM obtained from the analytical modelling, as learned from such simulations in the GR case where one finds $M \lesssim 0.08M_{\odot}$ \cite{kumar}), one could expect the viable range of (positive) $\alpha$ to be somewhat enhanced with respect to the results obtained here, though we point out again the quick growth of the $M_{MMSM}$ above $\alpha \gtrsim 0.01$. Let us also mention that, should one also have information on the central density and/or radius of that star, one could use it to obtain the constraint on the quadratic gravity parameter $\beta$ from $\alpha$. Another option would be to approximate the central density in our case by using Eq.(\ref{cored}) to get $ \rho_c \sim R^{-6}$.

\begin{table}[t!]
\begin{center}
\begin{tabular}{|c||c c|c c|c|}
\hline
$\alpha$ & $\xi_R$ & $\omega_{3/2}(\xi_R)$ & $\gamma_{3/2}(\xi_R)$  & $\delta_{3/2}(\xi_R)$ & $M/M\odot$  \\
\hline\hline
-0.100 & 3.64 & 2.39  & 2.25 & 6.67 & 0.0810 \\ 
-0.010 & 3.65 & 2.68  & 2.35 & 6.09 & 0.0910 \\ 
\hline
0 (GR)    & 3.65 & 2.71 &  2.36 & 5.97  & 0.0922 \\ 
\hline
0.006   & 3.66 & 2.73 & 2.36  & 5.95  & 0.0929 \\ 
0.010   & 3.66 & 2.75 & 2.37 & 5.93  & 0.0933\\  
0.015   & 3.66 & 2.77 &  2.46 & 5.89  & 0.0980\\
\hline
\end{tabular}
\caption{Numerical values of $\xi_R$ obtained from $\theta(\xi_R)=0$, and the associated values of the functions $\gamma_{3/2}$, $\omega_{3/2}$, and $\delta_{3/2}$  for different values of $\alpha=\kappa c^2 \beta \rho_c$. The last column provides the estimations for the normalized (in solar mass units) MMSM for each value of $\alpha$, which must be compared with the observational bound $0.0930\pm0.0008 M_\odot$ of the M-dwarf star G1 866C \cite{Segransan:2000jq}.}
\label{tab}
\end{center}
\end{table}

The above discussion and results show the feasibility of investigations of non-relativistic stars, illustrated here with the case of the MMSM of high-mass brown dwarfs, as astrophysical tests to constraint Palatini theories of gravity, in particular, the simple quadratic $f(\mathcal{R})$ model considered here. As already discussed, a major novelty introduced within these models lies on the fact that, due to the local contributions on the energy density that these theories introduce, the MMSM provides constraints upon a combination of the new gravitational parameter, $\alpha$, and the star's core density, $\rho_c$, where the latter have to be estimated by other means to obtain reliable bounds on the former.

To conclude, in combination with astrophysical tests of (both individual and merger of binaries) neutron stars and also with cosmological tests, more refined analyses of the MMSM could allow to narrow down the viable range of the parameters of these theories as compared with different observations.  Another path worth exploring in this context is the implementation, within astrophysical settings, of the recently found mapping between the Lagrangian densities and the spaces of solutions of GR and Ricci-based gravities (with $f(\mathcal{R})$ being just a particular case of the latter) for generic anisotropic fluids \cite{Afonso:2018bpv}, which might allow to circumvent some of the many limitations found in the analysis presented here. In particular, this mapping is expected to allow  for a direct implementation of the numerical codes developed within GR to the context described here for further analyses of the MMSM and other issues of the stellar structure modelling beyond GR. Finally, both the direct attack presented here and the shortcut provided by the mapping technique could also be used to study other Palatini theories of gravity beyond the $f(\mathcal{R})$ family, such as quadratic gravity with Ricci-squared terms, or Eddington-inspired Born-Infeld gravity and its many extensions \cite{BeltranJimenez:2017doy}.  Work along these lines is currently underway. \\

\acknowledgments{GJO is funded by the Ramon y Cajal contract RYC-2013-13019 (Spain). DRG is funded by the \emph{Atracci\'on de Talento Investigador} programme of the Comunidad de Madrid (Spain) No. 2018-T1/TIC-10431, and acknowledges further support from the Funda\c{c}\~ao para a Ci\^encia e a Tecnologia (FCT, Portugal) research projects Nos. UID/FIS/04434/2013, PTDC/FIS-OUT/29048/2017 and PTDC/FIS-PAR/31938/2017. AW acknowledges
financial support from FAPES (Brazil).
This work is supported by the Spanish projects FIS2014-57387-C3-1-P and FIS2017-84440-C2-1-P (MINECO/FEDER, EU), the project H2020-MSCA-RISE-2017
Grant FunFiCO-777740, the project SEJI/2017/042 (Generalitat Valenciana), the Consolider Program CPANPHY-1205388, the
Severo Ochoa grant SEV-2014-0398 (Spain), the Edital 006/2018 PRONEX (FAPESQ-PB/CNPQ, Brazil) and the EU COST Actions CA15117 and CA18108. DRG and AW thank the Department of Physics and IFIC of the University of Valencia for their hospitality during different stages of the elaboration of this work. }


\begin{thebibliography}{99}


  \bibitem{Will:2014kxa}
  C.~M.~Will,
  Living Rev.\ Rel.\  {\bf 17} (2014) 4.

\bibitem{TheLIGOScientific:2017first}
  B.~P.~Abbott {\it et al.} [LIGO Scientific and Virgo Collaborations],
Phys. Rev. Lett. \textbf{116} (2016) 061102.

\bibitem{TheLIGOScientific:2017qsa}
  B.~P.~Abbott {\it et al.} [LIGO Scientific and Virgo Collaborations],
  Phys.\ Rev.\ Lett.\  {\bf 119} (2017)  161101.



  \bibitem{Barack:2018yly}
  L.~Barack {\it et al.}, arXiv:1806.05195 [gr-qc].

\bibitem{Akiyama:2019cqa}
  K.~Akiyama {\it et al.} [Event Horizon Telescope Collaboration],  Astrophys.\ J.\  {\bf 875} (2019)  L1.



\bibitem{Copeland:2006wr}
E. J. Copeland, M.~Sami, and S.~Tsujikawa, Int. J. Mod. Phys. D \textbf{15} (2006) 1753.

\bibitem{Nojiri:2006ri}
S.~Nojiri and S. D. Odintsov, Int. J. Geom. Meth. Mod. Phys. \textbf{4}  (2007)  115.

\bibitem{Capozziello:2007ec}
S.~Capozziello and M.~Francaviglia, Gen. Rel. Grav. \textbf{40} (2008) 357.

\bibitem{Carroll:2004de}
S. M. Carroll, A.~De~Felice, V.~Duvvuri, D. A. Easson, M.~Trodden, and M. S. Turner, Phys. Rev. D \textbf{71} (2005) 063513.

\bibitem{ParTom}
L. Parker and D. J. Toms, \emph{``Quantum Field Theory in Curved Spacetime: Quantized Fields and Gravity"}
(Cambridge University Press, Cambridge, England, 2009).

\bibitem{BirDav}
N. D. Birrel and P. C. W. Davies, \emph{``Quantum Fields in Curved Space"} (Cambridge University Press, Cambridge, England, 1982).

\bibitem{Senovilla:2014gza}
  J.~M.~M.~Senovilla and D.~Garfinkle,  Class.\ Quant.\ Grav.\  {\bf 32} (2015)  124008.

  \bibitem{DeFelice:2010aj}
A.~De~Felice and S.~Tsujikawa, Living Rev.\ Rel. \textbf{13} (2010) 3.

\bibitem{brans} C. H. Brans and R. H. Dicke, Phys. Rev. \textbf{124} (1961) 925.

\bibitem{Bergmann} P. G. Bergmann, Int. J. Theor. Phys.  \textbf{1} (1968) 25.

\bibitem{BeltranJimenez:2019tjy}
  J.~Beltr\'an Jim\'enez, L.~Heisenberg and T.~S.~Koivisto,
  arXiv:1903.06830 [hep-th].

\bibitem{Dabrowski:2012eb}  M. P.~Dabrowski and K.~Marosek, JCAP  \textbf{1302} (2013)  012.

\bibitem{Leszczynska:2014xba} K.~Leszczynska, A.~Balcerzak and M. P.~Dabrowski, JCAP \textbf{1502} (2015) 012.

  \bibitem{Ezquiaga:2017ekz}
  J.~M.~Ezquiaga and M.~Zumalac\'arregui,
Phys.\ Rev.\ Lett.\  {\bf 119} (2017) 251304.

\bibitem{Baker:2017hug}
  T.~Baker, E.~Bellini, P.~G.~Ferreira, M.~Lagos, J.~Noller and I.~Sawicki,
  Phys.\ Rev.\ Lett.\  {\bf 119} (2017) 251301.
  
  \bibitem{Creminelli:2017sry}
  P.~Creminelli and F.~Vernizzi,
  Phys.\ Rev.\ Lett.\  {\bf 119} (2017) 251302.
  
  \bibitem{Langlois:2017dyl}
  D.~Langlois, R.~Saito, D.~Yamauchi and K.~Noui,
  Phys.\ Rev.\ D {\bf 97} (2018) 061501.
  
  \bibitem{Sakstein:2017xjx}
  J.~Sakstein and B.~Jain,
  Phys.\ Rev.\ Lett.\  {\bf 119} (2017)  251303.
  
  \bibitem{Lombriser:2015sxa}
  L.~Lombriser and A.~Taylor,
  JCAP {\bf 1603} (2016) 031.


\bibitem{Berti:2015itd}
  E.~Berti {\it et al.},  Class.\ Quant.\ Grav.\  {\bf 32} (2015) 243001.

\bibitem{lina} M. Linares, T. Shahbaz, and J. Casares, The Astrophysical Journal \textbf{859} (2018) 54.

\bibitem{as} J. Antoniadis {\it et al.}, Science \textbf{340} (2012)  6131.

\bibitem{craw} F. Crawford, M. S. E. Roberts, J. W. T. Hessels, S. M. Ransom, M. Livingstone, C. R. Tam and V. M. Kaspi, Astrophys. J. \textbf{652} (2006) 1499.

 \bibitem{Burrows:1992fg}
  A.~Burrows and J.~Liebert,    Rev.\ Mod.\ Phys.\  {\bf 65} (1993) 301.

\bibitem{Sakstein:2015zoa}
  J.~Sakstein,  Phys.\ Rev.\ Lett.\  {\bf 115} (2015) 201101.

\bibitem{sak} J. Sakstein, Phys. Rev. D \textbf{92} (2015) 124045.

\bibitem{Crisostomi:2019yfo}
  M.~Crisostomi, M.~Lewandowski and F.~Vernizzi,
  arXiv:1903.11591 [gr-qc].

\bibitem{Olmo:2011uz}
  G.~J.~Olmo,
  Int.\ J.\ Mod.\ Phys.\ D {\bf 20} (2011) 413.

\bibitem{Bambi:2015zch}
  C.~Bambi, A.~Cardenas-Avendano, G.~J.~Olmo and D.~Rubiera-Garcia, Phys.\ Rev.\ D {\bf 93} (2016) 064016.

\bibitem{Bejarano:2017fgz}
  C.~Bejarano, G.~J.~Olmo and D.~Rubiera-Garcia,  Phys.\ Rev.\ D {\bf 95} (2017) 064043.

\bibitem{Segransan:2000jq}
  D.~Segransan {\it et al.},  Astron.\ Astrophys.\  {\bf 364} (2000) 665.


\bibitem{Olmo:2005zr} 
  G.~J.~Olmo,
  Phys.\ Rev.\ Lett.\  {\bf 95}, 261102 (2005)
  doi:10.1103/PhysRevLett.95.261102
  [gr-qc/0505101].

\bibitem{Olmo:2005hc} 
  G.~J.~Olmo,
  Phys.\ Rev.\ D {\bf 72}, 083505 (2005)
  doi:10.1103/PhysRevD.72.083505
  [gr-qc/0505135].
  
\bibitem{Olmo:2011fh} 
  G.~J.~Olmo and H.~Sanchis-Alepuz,
  Phys.\ Rev.\ D {\bf 83}, 104036 (2011)
  doi:10.1103/PhysRevD.83.104036
  [arXiv:1101.3403 [gr-qc]].


\bibitem{BSS} A. Stachowski, M. Szydlowski, A. Borowiec, Eur. Phys. J. C \textbf{77} (2017) 406.

\bibitem{SSB} M. Szydlowski, A. Stachowski, A. Borowiec, Eur. Phys. J. C \textbf{77} (2017) 603.


\bibitem{anet} A. Wojnar, Eur. Phys. J C \textbf{78} (2018) 421.


\bibitem{aneta} A. Wojnar, H. Velten, Eur. Phys. J. C \textbf{76} (2016) 697.

 \bibitem{Starobinsky:1980te}
A. A. Starobinsky, Phys. Lett. B \textbf{91} (1980) 99.

\bibitem{Stelle:1977ry}
  K.~S.~Stelle,
  Gen.\ Rel.\ Grav.\  {\bf 9} (1978) 353.
  
  \bibitem{Deruelle:2009zk}
  N.~Deruelle, M.~Sasaki, Y.~Sendouda and D.~Yamauchi,
  Prog.\ Theor.\ Phys.\  {\bf 123} (2010) 169

\bibitem{Afonso:2017bxr} 
  V.~I.~Afonso, C.~Bejarano, J.~Beltran Jimenez, G.~J.~Olmo and E.~Orazi,
  Class.\ Quant.\ Grav.\  {\bf 34}, 235003 (2017).



\bibitem{Pani:2012qd}
  P.~Pani and T.~P.~Sotiriou,
  Phys.\ Rev.\ Lett.\  {\bf 109} (2012)  251102.

  \bibitem{Barausse:2007pn}
  E.~Barausse, T.~P.~Sotiriou and J.~C.~Miller,
  Class.\ Quant.\ Grav.\  {\bf 25} (2008) 062001.

  \bibitem{Olmo:2008pv}
  G.~J.~Olmo,
  Phys.\ Rev.\ D {\bf 78} (2008) 104026.

\bibitem{BeltranJimenez:2017doy}
  J.~Beltran Jimenez, L.~Heisenberg, G.~J.~Olmo and D.~Rubiera-Garcia,
  Phys.\ Rept.\  {\bf 727} (2018) 1.



\bibitem{aneta2} A. Wojnar, Eur. Phys. J C \textbf{79} (2019) 51.


\bibitem{Koyama:2015oma} 
  K.~Koyama and J.~Sakstein,
  Phys.\ Rev.\ D {\bf 91}, 124066 (2015).

\bibitem{Saito:2015fza} 
  R.~Saito, D.~Yamauchi, S.~Mizuno, J.~Gleyzes and D.~Langlois,
  JCAP {\bf 1506}, 008 (2015).


\bibitem{GlenBook}
N. K. Glendenning, \emph{``Compact Stars: Nuclear Physics, Particle Physics, and General Relativity"}, 2nd Edition (Astronomy and Astrophysics Library, Springer, 2000).


\bibitem{KippBook}
R. Kippenhahn, A. Weigert, A. Weiss, \emph{``Stellar Structure and Evolution"}, 2nd Edition  (Astronomy and Astrophysics Library, 
Springer, 2012).

\bibitem{Capozziello:2011nr}
  S.~Capozziello, M.~De Laurentis, S.~D.~Odintsov and A.~Stabile,
  Phys.\ Rev.\ D {\bf 83} (2011) 064004.


\bibitem{artur} A. Sergyeyev, A. Wojnar, arXiv:1901.10448 [gr-qc].

\bibitem{FowHoy}
W. A Fowler and F. Hoyle, Astrophys. J. Suppl. Ser. \textbf{9} (1964) 201.



\bibitem{Freedman:2007cm}
  R.~S.~Freedman, M.~S.~Marley and K.~Lodders,  Astrophys.\ J.\ Suppl.\  {\bf 174} (2008) 504.



\bibitem{kumar} S. S. Kumar, Astrophysics Journal \textbf{137} (1963) 1121.

\bibitem{Afonso:2018bpv}
  V.~I.~Afonso, G.~J.~Olmo and D.~Rubiera-Garcia, Phys.\ Rev.\ D {\bf 97} (2018)  021503.






\end{thebibliography}
\end{document}